\documentclass[conf]{IEEEtran}
\usepackage{amsmath,amssymb}
\usepackage{graphicx}
\usepackage{mathrsfs}
\usepackage{color}
\usepackage{tikz}
\usepackage{enumerate}
\usepackage{psfrag}	
\usepackage{array}
\usepackage{multirow,hhline}
\usepackage{exscale}
\usepackage{colortbl}
\usepackage{epsfig,subfigure}
\usepackage{cite}
\usepackage{amsthm}
\usepackage[bookmarks=false]{hyperref}
\usepackage{nicefrac}

\usepackage[outline]{contour}
\contourlength{1.2pt}
\usepackage{tikz}
\usepackage{pgfplots}
\usepackage{pgfplotstable}
\usepackage{rotate}
\usepgfplotslibrary{units}
\usetikzlibrary{%
patterns,%
calc,%
fit,%
arrows,%
plotmarks,%
shadows,%
chains,%
shapes%
}
\usepackage[outline]{contour}
\contourlength{1.2pt}
\tikzset{>=latex'} 
\tikzstyle{every picture}+=[remember picture] 
\definecolor{rubgray}{cmyk}{0.03,0.03,0.03,0.1}

\newcommand{\alert}[1]{\textcolor{black}{#1}}

\newtheorem{proposition}{Proposition}

\newcommand{\y}[0]{\ensuremath{\mathbf{y}}}

\newcommand{\z}[0]{\ensuremath{\mathbf{z}}}

\newcommand{\x}[0]{\ensuremath{\mathbf{x}}}

\newcommand{\snr}[0]{\ensuremath{\mathsf{SNR}}}

\begin{document}

\pagenumbering{arabic}
\pagestyle{plain}

\IEEEoverridecommandlockouts
\title{When Can a Relay Reduce End-to-End Communication Delay?}
\author{
\IEEEauthorblockN{Anas Chaaban and Aydin Sezgin}\\
\IEEEauthorblockA{Institute of Digital Communication Systems\\ 
Ruhr-Universit\"at Bochum (RUB), Germany}\\
Email: anas.chaaban@rub.de, aydin.sezgin@rub.de
}

\maketitle


\begin{abstract}
The impact of relaying on the latency of communication in a relay channel is studied. Both decode-forward (DF) and amplify-forward (AF) are considered, and are compared with the point-to-point (P2P) scheme which does not use the relay. The question as to whether DF and AF can decrease the latency of communicating a number of bits with a given reliability requirement is addressed. Latency expressions for the three schemes are derived. Although both DF and AF use a block-transmission structure which sends the information over multiple transmission blocks, they can both achieve latencies lower that P2P. Conditions under which this occurs are obtained. Interestingly, these conditions are more strict when compared to the conditions under which DF and AF achieve higher information-theoretic rates than P2P. 
\end{abstract}


\section{Introduction}
Low latency is an important requirement in many communications scenarios (security, emergency, etc.). In such scenarios, a number of bits has to be transmitted from a source to a destination within a given reliability (error probability) in the shortest time possible. Methods that reduce communication latency are thus of practical interest in such scenarios. Several methods have been examined for this purpose, such as channel coding \cite{PolyanskiyPoorVerdu} and feedback \cite{SchalkwijkKailath, Schalkwijk, KimLapidothWeissman, BurnashevYamamoto}. The question we examine in this paper is: Can a relay node be used to reduce latency?

The common way to look at this problem is by analysing the achievable information-theoretic rate in a network with relays. Indeed, relays can increase the communication rate \cite{CoverElgamal}. However, looking at the problem from this perspective has two practical draw-backs. First, the achievable rates are derived under the requirement that the error probability approaches zero as the number of transmission goes to infinity. Second, the resulting rate is achievable by transmitting over a large number of blocks and using backward decoding \cite{CoverElgamal,KramerGastparGupta,ChaabanSezgin_IT_IRC,
TianYener_PotentRelayJournal,MaricDaboraGoldsmith_IT}. Clearly this perspective is not suitable for low latency communications. 

Thus, this problem has to be approached from a different perspective. Namely, the latency has to be calculated for the given number of bits to be transmitted under the given error probability requirement. For a small but non-zero error probability requirement, the latency of communication is finite \cite{AgarwalGuoHonig, MaricLatency}. This means that for a point-to-point (P2P) channel  e.g., the latency is bounded. 

By installing a relay node into the channel, we get the so-called relay channel (RC). With the block structure of relaying schemes in mind, it might seem at first that a relay increases latency. However, we show in this paper that if the relay is properly placed, then the relay reduces the latency in the RC in comparison to the P2P channel. To do this, we derive the latency of two relaying schemes: decode-forward (DF) and amplify-forward (AF), and compare their latency to that of the P2P scheme which ignores the relay (benchmark). The latency is derived by making use of the error-exponent of Gaussian codes over an AWGN channel \cite{AgarwalGuoHonig}. We restrict ourselves to these three schemes (P2P, DF, AF) since their combination achieves the capacity of the RC within a constant gap \cite{AvestimehrDiggaviTse_IT}. 

The DF and AF schemes transmit the information over multiple blocks. Therefore, the payload (bits to be transmitted) have to be distributed over multiple blocks. In this case, the whole transmission will be erroneous if one block is erroneous. This imposes \alert{an error probability requirement per block $\epsilon'$ which is stricter than the error probability requirement for the whole transmission $\epsilon>\epsilon'$}. Consequently, this leads to long transmission blocks. The number of used transmission blocks has to be optimized for the given scenario. It turns out that for small payloads (number of bits), one transmission block is optimal, while multiple blocks yield better latency for a large payloads. Interestingly, in both cases the latency of DF and AF can still be lower than that of P2P.

For the high $\snr$ regime, conditions on the relay channels are derived (for DF and AF), under which relaying reduces latency, leading to the following conclusion. If DF or AF increase the capacity of the P2P channel, they do not necessarily reduces the latency of transmission. On the other hand, if either DF or AF reduces latency, then their achievable information-theoretic rate has to be higher than that of P2P. Therefore, for such problems, the information-theoretic achievable rate of a given scheme can be misleading. These aspects will be discussed in detail throughout the paper. In the next section, we introduce the system model of the RC and provide the problem formulation.

\section{System Model and Problem Formulation}
\label{Model}
\begin{figure}
\centering
\begin{tikzpicture}[semithick,scale=.8, every node/.style={scale=.8}]]
\node (t1) at (-1,0) [draw,thick,minimum width=.9cm,minimum height=.9cm] {Source};
\node (r1) at (6.5,0) [draw,thick,minimum width=.9cm,minimum height=.9cm] {Dest.};
\node (r) at (3,1.5) [draw,thick,minimum width=.9cm,minimum height=.9cm] {Relay};

\node (x1) at ($(t1.east)+(.7,0)$) {$x_s$};
\draw[->] ($(t1.east)+(0,0)$) to (x1.west);
\node (p1) at ($(r1.west)-(1.4,-0)$) [circle,draw, minimum width=.1cm,minimum height=.1cm] {};
\node at ($(r1.west)-(1.4,-0)$) {$+$};
\node (y1) at ($(r1.west)+(-.7,0)$) {$y_d$};
\draw[->] (y1.east) to ($(r1.west)+(0,0)$);
\draw[->] (p1.east) to (y1.west);
\draw[->] ($(p1.south)+(0,-.3)$) to ($(p1.south)$);
\node at ($(p1.south)+(0,-.5)$) {$z_d$};
\draw[->] (x1.east) to (p1.west);
\node at ($(x1.east)+(2,0)$) {\contour{white}{$h_0$}};

\node (m1) at ($(t1)-(1.3,0)$) {$m$};
\node (m1h) at ($(r1)+(1.2,0)$) {$\hat{m}$};
\draw[->] (m1.east) to (t1.west);
\draw[->] (r1.east) to (m1h.west);

\node (pr) at ($(r.west)-(1.4,-0)$) [circle,draw, minimum width=.1cm,minimum height=.1cm] {};
\node at ($(r.west)-(1.4,-0)$) {$+$};
\node (yr) at ($(r.west)+(-.7,0)$) {$y_r$};
\draw[->] (yr.east) to ($(r.west)+(0,0)$);
\draw[->] (pr.east) to (yr.west);
\draw[->] ($(pr.west)+(-.3,0)$) to ($(pr.west)$);
\node at ($(pr.west)+(-.5,0)$) {$z_r$};
\draw[->] ($(pr)-(0,1.5)$) to (pr.south);
\node at ($(pr)-(0,.75)$) {\contour{white}{$h_1$}};

\node (xr) at ($(r.east)+(1.05,0)$) {$x_r$};
\draw[->] ($(r.east)+(0,0)$) to (xr.west);
\draw[->] (xr.south) to (p1.north);
\node at ($(xr)+(0,-.75)$) {\contour{white}{$h_2$}};
\end{tikzpicture}
\caption{A mathematical model of the Gaussian relay channel.}
\label{Fig:Model}
\end{figure}
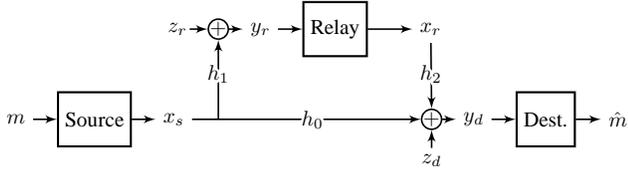

We consider a relay channel (RC) as shown in Fig. \ref{Fig:Model} where the source node wants to send a message $m$ of $B$ bits to the destination node with the aid of a full-duplex relay. At time instant $i\in\{1,\cdots,N\}$, the source sends the real-valued signal $x_s(i)$ to the relay and the destination. These nodes in turn observe the received signals,
\begin{align}
y_r(i)&=h_1x_s(i)+z_r(i),\\
y_d(i)&=h_0x_s(i)+h_2x_r(i)+z_d(i),
\end{align}
respectively. Here, $x_r(i)$ is the relay transmit signal constructed in general from $y_r(1),\cdots,y_r(i-1)$. The variables $z_r(i)$ and $z_d(i)$ are independent Gaussian noises distributed as $\mathcal{N}(0,1)$, and $h_0,h_1,h_2\in\mathbb{R}$ are the source-destination, source-relay, and relay-destination channel coefficients, respectively. It is assumed that these channel coefficients maintain the same value for the whole transmission duration. Each of the source and the relay have a power constraints given by $\mathbb{E}[x_s^2]\leq P$, $\mathbb{E}[x_r^2]\leq P_r$. After $N$ transmissions, the destination decodes $\hat{m}$ from $y_d(1),\cdots,y_d(N)$. The induced error probability of this procedure $P_e=\mathbb{P}\{m\neq \hat{m}\}$ has to satisfy
\begin{align}
\label{PeReq}
P_e< \epsilon,
\end{align} 
where $\epsilon>0$ is a pre-defined reliability requirement.

The main goal of this work is to study the latency of this communication, and whether the relay has a positive impact on latency. More precisely, we aim for finding the value of $N$ that has to be chosen so that $B$ bits can be delivered to the destination with an error probability satisfying \eqref{PeReq}. 

An important quantity for this study is the error exponent of a coding scheme defined as \cite{XiangKim}
\begin{align}
E_r=-\lim_{N\to\infty} (\nicefrac{1}{N})\ln(P_e).
\end{align}
Gaussian coding with rate $R$ over a Gaussian P2P channel with signal-to-noise ratio $\mathsf{SNR}$ achieves an error exponent~\cite{AgarwalGuoHonig}
\begin{align}
E_r(R,\mathsf{SNR})\geq\max_{\rho\in[0,1]}\left[\frac{\rho}{2}\log_2\left(1+\frac{\mathsf{SNR}}{1+\rho}\right)-\rho R\right].
\end{align}
Given $E_r(R,\snr)$, the latency of communicating $B$ bits over this channel with reliability $\epsilon$ is upper bounded by
\begin{align}
\label{LatencyFunction}
n(B,\mathsf{SNR},\epsilon)=\min_{\rho\in[0,1]}\frac{\rho B-\ln(\epsilon)}{\frac{\rho}{2}\log_2\left(1+\frac{\mathsf{SNR}}{1+\rho}\right)}.
\end{align}
This function $n(B,\mathsf{SNR},\epsilon)$ will be used frequently in the paper for bounding the latency of some transmission schemes over the RC. Next, we summarize the main contribution of the paper.

\section{Summary of Main Results}
In the following sections, we will derive the achievable latency of using Gaussian coding in the RC with DF and AF at the relay. We are going to prove that DF and AF achieve a latency of 
\begin{align}
\label{Eq:NDFfinal}
&\hspace{-.2cm}N_{DF}=\hspace{-.2cm}\min_{(L,\delta)\in\mathcal{L}}\hspace{-.2cm}\max\left\{[1\ L]\begin{bmatrix}n_1(L,\delta)\\n_2(L,\delta)\end{bmatrix}, [L\ 1]\begin{bmatrix}n_1(L,\delta)\\n_2(L,\delta)\end{bmatrix}\right\},
\end{align}
where $\mathcal{L}=[\mathbb{N}\setminus\{0\}]\times (0,1)$, and
\begin{align}
\label{Eq:NAFfinal}
N_{AF}&=\min_{L\in\mathbb{N}\setminus\{0\}} (L+1)\cdot n_3(L),
\end{align}
respectively, where
\begin{align}
\label{Eq:n1}
n_1(L,\delta)&= n(B/L,\snr_{DF1},(1-\delta)\epsilon/L)\\
\label{Eq:n2}
n_2(L,\delta)&= n(B/L,\snr_{DF2},\delta\epsilon/L)\\
\label{Eq:n3}
n_3(L)&= n(B/L,\snr_{AF},\epsilon/L),
\end{align}
with $\snr_{DF1}=h_1^2P$, $\snr_{DF2}=h_2^2P_r$, and $\snr_{AF}=\frac{h_1^2h_2^2PP_r}{1+h_1^2P+h_2^2P_r}$. Here, the parameter $L$ is the number of transmission blocks and $\delta$ is a trade-off factor \alert{which allows different error probabilities and block-lengths in the uplink and downlink of DF}. Notice the strict error probability requirement represented by $\epsilon/L$ arising due to the block structure. 

\begin{figure}[t]
\centering
\psfragscanon
\psfrag{y}[t]{$y$}
\psfrag{x}[t]{$x$}
\includegraphics[width=\columnwidth]{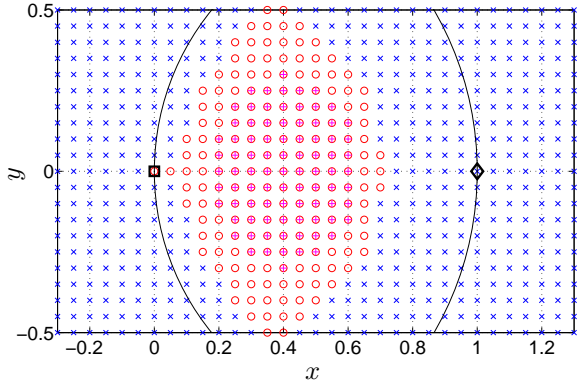}
\caption{The best scheme versus the relay position for an RC with a path-loss exponent of $3$, $P=20$dB, $P_r=2P$, $\epsilon=10^{-3}$, and $B=10$kbit. The $\Box$, $\Diamond$, $\mathsf{x}$, $\mathsf{o}$, and $\mathsf{+}$ denote the source, the destination, relay positions where $N_{P2P}<N_{DF}<N_{AF}$, $N_{DF}<N_{P2P}<N_{AF}$, and $N_{DF}<N_{AF}<N_{P2P}$, respectively. \alert{The area between the two black curves passing through the source and destination is the region where $R_{DF}>R_{P2P}$.}}
\label{Fig:Rpos}
\end{figure}

DF and AF achieve lower latency than P2P which achieves
\begin{align}
\label{P2PLatency}
N_{P2P}= n(B,\snr_{P2P},\epsilon)
\end{align}
where $\snr_{P2P}=h_0^2P$, if the relay is properly placed (see Figure \ref{Fig:Rpos}). \alert{In this figure, we show the best scheme in terms of latency versus the relay position. The source and destination are located at positions $(0,0)$ and $(1,0)$, respectively (normalized units of distance). We assume that the wireless channel has a path-loss exponent of 3. The channels $h_1$ and $h_2$ depend on the relay position. If the relay is located in the region marked by $\mathsf{x}$'s, then P2P achieves lower latency than both DF and AF. However, if the relay is located in the region marked by $\mathsf{o}$'s, then DF achieves lower latency than both AF and P2P. These positions marked by $\mathsf{o}$ are potential positions where a relay might be placed in a cellular communications scenario for instance, since the relay is normally located between the transmitter and the receiver. Note that the region marked with $\mathsf{o}$'s is a sub-set of the region bounded between the two black curves. This region indicates positions where DF achieves higher information-theoretic rate\footnote{Throughout the paper, we use 'information-theoretic rate' to refer to the achievable rate under the condition that $P_e\to 0$ as $N\to\infty$.} given by 
\begin{align}
\label{RateDF}
R_{DF}\approx\frac{1}{2}\log(\min\{\snr_{DF1},\snr_{DF2}\})
\end{align}
than P2P which achieves 
\begin{align}
\label{RateP2P}
R_{P2P}\approx\frac{1}{2}\log(\snr_{P2P}).
\end{align}
This interestingly means that if DF increases the achievable rate, it does not necessarily reduce latency. However, in the inner region marked by $\mathsf{o}$'s, DF indeed reduces latency in comparison to P2P.}

\alert{While DF provides lower latency than AF, the latter has the advantage of reduced computational requirements at the relay node. Thus, in cases where the relay has computational limitation, AF can be a favoured scheme in the region marked by $\mathsf{+}$'s where AF reduces latency in comparison with P2P.}

To obtain a closer look on the conditions under which a relay reduces latency in a RC, we consider the high $\snr$ regime where we have the following statement regarding DF.
\begin{proposition}
\label{Prop:DF}
At high $\snr$, if 
\begin{align}
\label{DFCond1}
\frac{1}{\log_2(\snr_{DF1})}+\frac{1}{\log_2(\snr_{DF2})}<\frac{1}{\log_2(\snr_{P2P})},
\end{align}
then DF has a lower latency than P2P.
\end{proposition}
\alert{This statement quantifies the observation in Fig. \ref{Fig:Rpos}: If DF increases the information-theoretic rate with respect to P2P, it does not necessarily reduce latency!} Note that while the information-theoretic rate of DF is dictated by the bottle-neck $\snr$ between $\snr_{DF1}$ and $\snr_{DF2}$ \eqref{RateDF}, the latency is determined by both $\snr$'s \eqref{DFCond1}. A similar statement holds for AF, for which we have the following statement.
\begin{proposition}
\label{Prop:AF}
At high $\snr$, if 
\begin{align}
\label{AFCond1}
\frac{2}{\log_2(\snr_{AF})}<\frac{1}{\log_2(\snr_{P2P})},
\end{align}
then AF has a lower latency than P2P.
\end{proposition}
Both the DF and AF schemes can reduce the latency of transmission, but under a stricter condition than merely having a larger information-theoretic rate. 

It turns out that in general, transmission using DF or AF should be carried out over only one transmission block for a small payload, but over several blocks for large payload. Interestingly, although the use of multiple transmission blocks imposes a stricter reliability requirement per block, the overall transmission can still have lower latency than P2P. Next, we describe the three main transmission schemes of this paper.

\section{Transmission Schemes and their Latency}
\label{Sec:Schemes}
The number of transmissions required to satisfy \eqref{PeReq} depends on the scheme being used over the RC. The benchmark for our work is the scheme without a relay. The reason to choose this scheme as a benchmark is to check if the relay can in fact decrease the latency of this communication. If the relay is inactive, then the RC becomes a point-to-point channel (P2P) with $\snr_{P2P}=h_0^2P$. The optimal code for this P2P channel is a random Gaussian code \cite{CoverThomas}. In this case, the source encodes the message $m$ of $B$ bits into a sequence of length $N_{P2P}$ whose components are i.i.d. $\mathcal{N}(0,P)$. The destination decodes after observing $N_{P2P}$ received symbols. The latency of this scheme is given by $N_{P2P}$, which has to be chosen such that the error probability is below $\epsilon$. \alert{By using \eqref{LatencyFunction}, the $B$ bits can be delivered in this case with reliability $\epsilon$ if $N_{P2P}$ is chosen such that \eqref{P2PLatency} is satisfied. Next, we describe schemes that incorporate the relay.}

\subsection{Decode-forward}
In decode-forward (DF), the relay decodes the signal sent by the source, and forwards it to the destination in the next transmission block. We use here a simple variant of DF which does not incorporate superposition block-Markov encoding \cite{CoverElgamal, SahinErkip}. This simplification is made since the channel capacity can be achieved within a constant gap by DF without superposition block-Markov encoding \cite{AvestimehrDiggaviTse_IT}. Furthermore, this simplifies the analysis of the problem at hand. 

\subsubsection{Encoding-decoding}
The source splits $m$ into $L$ equal parts, denoted $m_1,\cdots,m_L$, each with $B'=B/L$ bits each. Then, the source encodes each message $m_\ell$, $\ell=1,\cdots,L$, into a codeword $\x_{s,\ell}$ of length $N_1$ using a Gaussian code with power $P$. Afterwards, the source sends $\x_{s,\ell}$ in the $\ell$-th transmission block. 

The relay waits until it has received $N_1$ symbols, after which it decodes $m_{r,1}$ (which is equal to $m_1$ unless an error occurs). Thus, the channel from the source to the relay is treated as a P2P channel with signal-to-noise ratio $\snr_{DF1}=h_1^2P$, leading to the first term in \eqref{RateDF}. The relay then encodes $m_{r,1}$ into $\x_{r,1}$ using a Gaussian code with power $P_r$ and length $N_2$, and sends $\x_{r,1}$ in the first relaying block. The first relaying block begins at time instant $\tau+1$ where $\tau\geq N_1$ to be determined later. The relay proceeds similarly by decoding $m_{r,\ell}$ after the end of the $\ell$-the transmission block, and forwarding it in the $\ell$-th relaying block\footnote{\alert{In this paper, we assume that the propagation delay and the decoding delay at the destination are negligible compared to the transmission delay.}}, until all message parts have been sent. The whole process takes $N_{DF}=\tau+L N_2$.

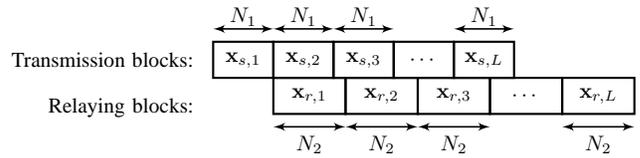
\begin{figure}[t]
\centering
\begin{tikzpicture}[semithick,scale=.8, every node/.style={scale=.8}]]
\node (tb1) at (0,0) [draw,thick,minimum width=1cm,minimum height=.6cm] {$\x_{s,1}$};
\node (tb2) at (1,0) [draw,thick,minimum width=1cm,minimum height=.6cm] {$\x_{s,2}$};
\node (tb3) at (2,0) [draw,thick,minimum width=1cm,minimum height=.6cm] {$\x_{s,3}$};
\node at (3,0) [draw,thick,minimum width=1cm,minimum height=.6cm] {$\cdots$};
\node (tbl) at (4,0) [draw,thick,minimum width=1cm,minimum height=.6cm] {$\x_{s,L}$};

\node (rb1) at (1.1,-.6) [draw,thick,minimum width=1.2cm,minimum height=.6cm] {$\x_{r,1}$};
\node (rb2) at (2.3,-.6) [draw,thick,minimum width=1.2cm,minimum height=.6cm] {$\x_{r,2}$};
\node (rb3) at (3.5,-.6) [draw,thick,minimum width=1.2cm,minimum height=.6cm] {$\x_{r,3}$};
\node at (4.7,-.6) [draw,thick,minimum width=1.2cm,minimum height=.6cm] {$\cdots$};
\node (rbl) at (5.9,-.6) [draw,thick,minimum width=1.2cm,minimum height=.6cm] {$\x_{r,L}$};

\draw[<->] ($(tb1.north)+(-.5,.2)$) to ($(tb1.north)+(.5,.2)$);
\node at ($(tb1.north)+(0,.4)$) {$N_1$};
\draw[<->] ($(tb2.north)+(-.5,.2)$) to ($(tb2.north)+(.5,.2)$);
\node at ($(tb2.north)+(0,.4)$) {$N_1$};
\draw[<->] ($(tb3.north)+(-.5,.2)$) to ($(tb3.north)+(.5,.2)$);
\node at ($(tb3.north)+(0,.4)$) {$N_1$};
\draw[<->] ($(tbl.north)+(-.5,.2)$) to ($(tbl.north)+(.5,.2)$);
\node at ($(tbl.north)+(0,.4)$) {$N_1$};

\draw[<->] ($(rb1.south)+(-.6,-.2)$) to ($(rb1.south)+(.6,-.2)$);
\node at ($(rb1.south)+(0,-.5)$) {$N_2$};
\draw[<->] ($(rb2.south)+(-.6,-.2)$) to ($(rb2.south)+(.6,-.2)$);
\node at ($(rb2.south)+(0,-.5)$) {$N_2$};
\draw[<->] ($(rb3.south)+(-.6,-.2)$) to ($(rb3.south)+(.6,-.2)$);
\node at ($(rb3.south)+(0,-.5)$) {$N_2$};
\draw[<->] ($(rbl.south)+(-.6,-.2)$) to ($(rbl.south)+(.6,-.2)$);
\node at ($(rbl.south)+(0,-.5)$) {$N_2$};

\node[left] at ($(tb1)-(.7,0)$) {Transmission blocks:};
\node[left] at ($(tb1)-(.7,.8)$) {Relaying blocks:};
\end{tikzpicture}
\caption{The block structure of the transmission using DF at the relay for the case $N_1<N_2$. For $N_1>N_2$, the last relaying block starts at the end of the last transmission block.}
\label{Fig:BlockStruct}
\end{figure}

During the whole transmission and relaying time, the destination simply listens and stores the received signals. At the end of the  transmission, the destination starts decoding backwards. The destination starts by decoding the last message part $\hat{m}_{r,L}$. We require that that $\x_{r,L}$ is received free of interference (from $\x_{s,L}$) at the relay. \alert{This is achieved by ensuring that the transmission of $\x_{s,L}$ from the source is completed before the transmission of $\x_{r,L}$ from the relay starts, as shown in Fig. \ref{Fig:BlockStruct}.} Hence, $L N_1\leq\tau+(L-1)N_2$ which yields $\tau\geq LN_1-(L-1)N_2$. Thus, by choosing $\tau=\max\{N_1,LN_1-(L-1)N_2\}$  (see Fig. \ref{Fig:BlockStruct}), the overall duration of communication in DFB becomes $$N_{DF}=\max\{N_1+LN_2,LN_1+N_2\}.$$
By using this procedure, the channel from the relay to the destination becomes a P2P channel with signal-to-noise ratio $\snr_{DF2}=h_2^2P_r$ leading to the second term in \eqref{RateDF}. Assume that $\hat{m}_{r,L}=m_L$. In this case, the destination constructs the codeword $\x_{s,L}$ and uses it to cancel the contribution of $\x_{s,L}$ from its received signal. Then the destination decodes the last but one message part $\hat{m}_{r,L-1}$, and proceeds similarly until all message parts have been recovered at the destination.

Notice that if $\hat{m}_{r,L}\neq m_L$, then perfect interference cancellation of $\x_{s,L-1}$ can not be carried out. In this case, an error might occur while decoding $\hat{m}_{L-1}$. This error propagates till block $\ell=1$. To calculate the latency of DF, we have to calculate the error probability of this scheme.

\subsubsection{Error probability and latency}
An error occurs in DF if an error occurs in block $L$ due to error propagation\footnote{This is a worst case consideration since an error in block $L$ might, but does not necessarily, lead to an error in block $L-1$.} resulting from the user of backward decoding and interference cancellation. An error occurs in block $L$ if the event $E_L=\{\hat{m}_{r,L}\neq m_L\}$ occurs. Let us call the probability of this event $P_{e,b}$. If no error occurs in block $L$, i.e., the event $\overline{E}_L$ (negation of $E_L$) occurs, then interference cancellation of $\x_{s,L}$ works. In this case, an error occurs if the event $E_{L-1}=\{\hat{m}_{r,L-1}\neq m_{L-1}\}$ occurs. The event $E_{L-1}$ occurs with a probability $P_{e,b}$ since all blocks are treated similarly, and the channel does not change between the blocks. By arguing similarly, we can write the error probability of DF as follows
\begin{align*}
P_{e,DF}&=\sum_{\ell=1}^L\mathbb{P}\{E_{\ell}|\overline{E}_{\ell+1},\cdots,\overline{E}_{L}\}\leq\sum_{\ell=1}^L\mathbb{P}\{E_{\ell}\}= L P_{e,b}.
\end{align*}
Therefore, if we can guarantee that $P_{e,b}<\frac{\epsilon}{L}=\epsilon'$, then we guarantee that $P_{e,DF}<\epsilon$. 

We conclude that we can only tolerate a block error probability $P_{e,b}<\epsilon'$. However, when does a block error occur in DF? A block error occurs if $m_\ell\neq \hat{m}_{r,\ell}$. This in turn occurs if either relay decodes correctly while the destination does not, or if both the relay and the destination decode the received signal incorrectly. Thus, $P_{e,b}$ can be upper bounded by the probability of error at the destination plus the probability of error at the relay, i.e., $P_{e,b}<P_{e,d}+P_{e,r}$. If we set $P_{e,d}<\delta\epsilon'$ and $P_{e,r}<(1-\delta)\epsilon'$, $\delta\in(0,1)$, then we guarantee that $P_{e,b}<\epsilon'$. The advantage of this trade-off parameter $\delta$ is to exploit the better channel among $h_1$ and $h_2$ to allow higher error probability tolerance at the weaker channel. 

From this calculation, we conclude that by increasing $L$ we decrease the number of bits $B'$ that should be delivered per block, thus decreasing $N_1$ and $N_2$. In return, we get more blocks. Additionally, we get a stricter reliability requirement per block given by $P_{e,b}<\epsilon'<\epsilon$ which is further reduced to $\delta\epsilon'$ and $(1-\delta)\epsilon'$ due to decoding each message twice. This in turn increases the block length $N_1$ and $N_2$. Whether this increase in $L$ has an advantage strongly depends on the parameters of the system $B$, $\epsilon$, $\snr_{DF1}$, and $\snr_{DF2}$.

Having bounded $P_{e,r}$ and $P_{e,d}$, now we can bound the length of blocks $N_1$ and $N_2$ using \eqref{LatencyFunction} by $N_1\geq n_1(L,\delta)$ and $N_2\geq n_2(L,\delta)$ as given in \eqref{Eq:n1} and \eqref{Eq:n2}. The parameters $L$ and $\delta$ has to be chosen so that the latency of DF $N_{DF}$ is minimized. Therefore, the minimum latency of DF can thus be written as given in \eqref{Eq:NDFfinal}.

\subsection{Amplify-forward}
Note that DF requires guaranteeing a reliability requirement not only at the destination, but also at the relay. The reliability requirement at the relay can be avoided by refraining from decoding at the relay and using amplify-forward instead.

\subsubsection{Encoding-decoding}
In AF, the source encodes similar to DF, by splitting $m$ into $L$ messages ($m_1,\cdots,m_L$) and sending the messages in $L$ transmission blocks. We denote the length of the codeword used by the source by $N_3$. Similar to DF, the relay waits until time instant $\tau\geq N_3$, and then starts transmission at time instant $\tau+1$. 

The relay scales the received signal $y_r(i)$ at time instant $i$ through multiplying by $\alpha=\sqrt{\nicefrac{P_r}{1+h_1^2P}}$ and sends it in time instant $i+\tau$. This scaling guarantees the satisfaction of the power constraint at the relay. This AF scheme leads to an equal length of transmission and relaying blocks, i.e., the length of the relaying block is also $N_3$. 

The destination receives a noisy superposition of the transmit and relay signals. It starts decoding from the last block. To guarantee that the last relaying block is free of interference, we need to choose $\tau=N_3$. Thus, the destination receives 
\begin{align}
\label{ydL}
\y_{d,L}=h_2\alpha(h_1\x_{s,L}+\z_{r,L})+\z_{d,L},
\end{align}
in the $L$-th relaying block, where $\x_{s,L}$ is the source signal corresponding to $m_L$, $\z_{r,L}=(z_r([L-1]N_3+1),\cdots,z_r(LN_3))$ is the additive noise at the relay during the source's $L$-th transmission block, and $\z_{d,L}=(z_d(LN_3+1),\cdots,z_d([L+1]N_3))$ is the additive noise at the destination during the $L$-th relaying block. The destination decodes $\hat{m}_L$ from \eqref{ydL} which resembles a P2P channel with signal-to-noise ratio of $$\snr_{AF}=\frac{h_2^2h_1^2PP_r}{1+h_1^2P+h_2^2P_r}.$$ Assuming $\hat{m}_L=m_L$ is decoded correctly, it is used by the destination to cancel the contribution of $\x_{s,L}$ from the $(L-1)$-th relaying block. Next, the destination decodes $\hat{m}_{L-1}$. This proceeds until all message parts are decoded. 

\subsubsection{Error probability and latency}
Since AF does not require decoding at the relay, this relaxes the error probability requirement since we do not need the parameter $\delta$ anymore. However, this comes at the expense of a reduced $\snr$. Following similar arguments as for the error probability of DF, we can write the error probability requirement per block of AF as $P_{e,b}<\epsilon'=\epsilon/L$. Since AF has an $\snr$ of $\snr_{AF}$, and since we need to send $B'=B/L$ bits per block, the block size of AF can be written as $N_3\geq n_3(L)$ where $n_3$ is defined in \eqref{Eq:n3} leading to the latency given in \eqref{Eq:NAFfinal}.

%

%
%
%
%
%
%
%
%

\subsection{Comparison}
\begin{figure*}[t]
\centering
\subfigure[$B=1$kbit.]{
\psfragscanon
\psfrag{x}[t]{\footnotesize Latency (transmissions)}
\psfrag{y}[b]{\footnotesize Reliability requirement $\epsilon$}
\psfrag{P2P}[l]{\footnotesize P2P}
\psfrag{DF}[l]{\footnotesize DF}
\psfrag{AF}[l]{\footnotesize AF}
\includegraphics[width=.95\columnwidth]{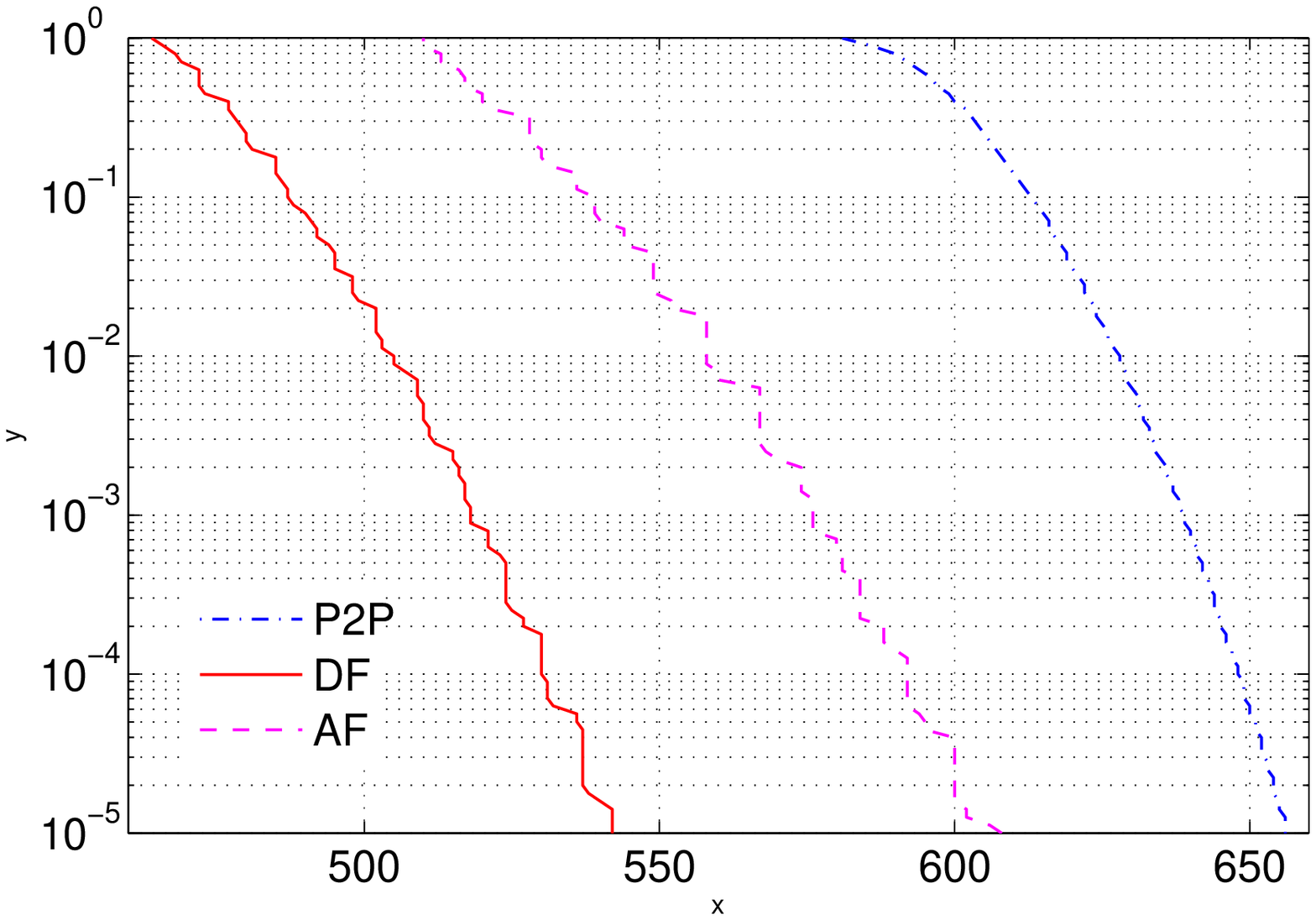}
\label{Fig:Nfe}
}
\hspace{.3cm}
\subfigure[$\epsilon=10^{-3}$]{
\psfragscanon
\psfrag{y}[b]{\footnotesize Latency (transmissions)}
\psfrag{x}[t]{\footnotesize $B$ (bits)}
\psfrag{P2P}[l]{\footnotesize P2P}
\psfrag{DF}[l]{\footnotesize DF}
\psfrag{AF}[l]{\footnotesize AF}
\includegraphics[width=.95\columnwidth]{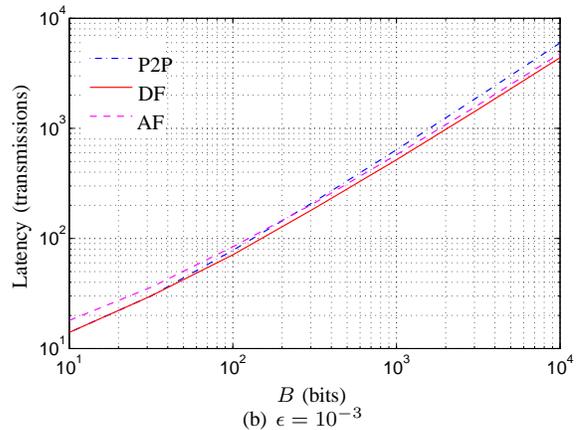}
\label{Fig:NfB}
}
\caption{The performance of the P2P, DF, and AF schemes for a relay channel with $h_0=h_2=1$, $h_1=2$, $P=10$dB, $P_r=16P$.}
\label{Fig:Performance}
\end{figure*}

It is intuitively clear that if $\snr_{P2P}\gg \snr_{DF1},\snr_{DF2}$, then the latency of the P2P is smaller than that of the DF scheme. Similar statement holds for AF. That is, if the channels to and from the relay are weak, then the relay has a negative impact on the latency of the communication. However, if the relay channels are strong enough, then a reduction of the latency can be achieved by relaying. This can be seen in Figure \ref{Fig:Nfe} which shows the latency of each of the P2P, DF, and AF schemes at different reliability requirements, for an RC with $h_0=h_2=1$, $h_1=2$, and $P=10$dB. This setting models a scenario where the relay is close to the source for instance, and the source and relay are equidistant from the destination. Furthermore, it is assumed that $P_r=16P$ which models scenarios where the relay is e.g. a fixed device which is mounted on a building having access to abundant power ($12$dB more than the source). Among the 3 schemes considered in this figure, the best in this case is DF. However, if less computational complexity is required at the relay, then AF can also be used to reduce the latency of the communication.


In Figure \ref{Fig:NfB}, the latency is plotted as a function of the message size $B$. In this figure, we can see that the performance of DF is close to that of P2P at small $B$. Recall that the block structure of DF has an advantage and a disadvantage. The advantage is the decreased number of bits to be delivered per block. The disadvantage is that these bits have to be delivered will a lower error probability. At low $B$, the advantage is lost since $B$ becomes negligible in comparison to $\ln(\epsilon)$. In other words, the function $n(B,\snr,\epsilon)$ approaches $n(0,\snr,\epsilon)$ at low $B$, and thus, at low $B$, dividing the number of bits to delivered per block by $L$ is irrelevant. This explains the behaviour of DF at low $B$ in Figure \ref{Fig:NfB}. However, at high $B$, this advantage becomes prominent, and DF becomes better than P2P. In this example, at $B=10$kbit we have a decrease in latency from $\approx 6000$ transmissions for P2P to $\approx 4400$ transmissions for DF, a drop of $>25\%$. 

Next, we analyse the performance of the three schemes described in Section \ref{Sec:Schemes} at high $\snr$ in order to obtain the statements of Propositions \ref{Prop:DF} and \ref{Prop:AF}.

\section{High $\snr$ Analysis}
\label{Sec:HighSNR}
We start by approximation the latency of the P2P scheme at high $\snr$.

\subsection{Latency of the P2P scheme}
The function $n(B,\snr,\epsilon)$ can be approximated at high $\snr$ as follows 
\begin{align*}
n(B,\mathsf{SNR},\epsilon)
&\approx\min_{\rho\in[0,1]}\frac{\rho B-\ln(\epsilon)}{\frac{\rho}{2}\log_2(\mathsf{SNR})}=\frac{2B-2\ln(\epsilon)}{\log_2(\mathsf{SNR})}.
\end{align*}
We use this approximation to write the latency of the P2P scheme at high $\snr$ as
\begin{align}
N_{P2P}\approx\frac{2B-2\ln(\epsilon)}{\log_2(\snr_{P2P})}.
\end{align}
Next, we use the approximation of $n(B,\snr,\epsilon)$ to express the latency of DF and AF at high $\snr$. To this end, we consider large $P$ and $P_r$ leading to large $\snr_{DF1}$, $\snr_{DF2}$, and $\snr_{AF}$.

\subsection{Latency of the DF scheme}
At high $\snr$, the block length parameters of the DF scheme can be approximated as
\begin{align}
n_1(L,\delta)&\approx\frac{2\frac{B}{L}-2\ln((1-\delta)\frac{\epsilon}{L})}{\log_2(\snr_{DF1})},
\\
n_2(L,\delta)&\approx\frac{2\frac{B}{L}-2\ln(\delta\frac{\epsilon}{L})}{\log_2(\snr_{DF2})}
.
\end{align}
To approximate the latency of DF, we need to minimize $\max\{n_1(L,\delta)+Ln_2(L,\delta),Ln_1(L,\delta)+n_2(L,\delta)\}$ over $L\in\mathbb{N}\setminus\{0\}$ and over $\delta\in(0,1)$. But before we proceed, this is a good point to discuss the impact of $L$ on latency. Let us examine the behaviour of $f_1(L)=n_1(L,\delta)+Ln_2(L,\delta)$ as a function of $L$. One can easily verify that 
the derivative of $f_1(L)$ by $dL$ is negative for small $L$ and positive for large $L$ if $B$ is large enough, and that this derivative is always positive if $B$ is small. A similar behaviour holds for $f_2(L)=Ln_1(L,\delta)+n_2(L,\delta)$. This leads to the following interesting conclusion. If $B$ is large enough, then the optimum $L$ is larger than 1. Consequently, high $\snr$ and high $B$, it is best to divide $B$ into several blocks to minimize latency. On the other hand, for small $B$, choosing $L=1$ is optimal. 

Instead of minimizing the latency with respect to $L$ and $\delta$, we bound $N_{DF}$ by choosing $\delta=\frac{1}{2}$ and $L=1$ to obtain
\begin{align*}
N_{DF}\leq \left[2B-2\ln\left(\frac{\epsilon}{2}\right)\right]\left[\frac{1}{\log_2(\snr_{DF1})}+\frac{1}{\log_2(\snr_{DF2})}\right]
\end{align*}

\subsection{Latency of the AF scheme}
At high $P$ and $P_r$, $\snr_{AF}$ is also high. The block length of the AF scheme is given by $N_3\geq n_3(L)$ where 
\begin{align}
n_3(L)\approx\frac{2\frac{B}{L}-2\ln\left(\frac{\epsilon}{L}\right)}{\log_2(\mathsf{SNR}_{AF})}.
\end{align}
The total latency of the AF scheme is thus given by
\begin{align}
N_{AF}=(L+1)\cdot n_3(L)\approx(L+1)\frac{2\frac{B}{L}-2\ln\left(\frac{\epsilon}{L}\right)}{\log_2(\mathsf{SNR}_{AF})}.
\end{align}
The behaviour of $N_{AF}$ as a function of $L$ is similar to $f_1(L)$, in the sense that it is decreasing and then increasing for large $B$, and only increasing for small $B$. Thus, the optimal $L$ is 1 for small $B$ and larger than 1 for larger $B$. We can bound $N_{AF}$ by setting $L=1$ as follows
\begin{align}
N_{AF}\leq\frac{4B-4\ln\left(\epsilon\right)}{\log_2(\mathsf{SNR}_{AF})}.
\end{align}

\subsection{Comparison}

Although we have set $L=1$ to upper bound the latency of DF and AF, the resulting latency upper bound of both schemes can be lower than that of P2P at high $\snr$. To show this, we compare $N_{P2P}$ and the upper bound for $N_{DF}$, to obtain the statement of Proposition \ref{Prop:DF}. Namely, at high $\snr$, if 
\begin{align}
\label{DFCond1}
\frac{1}{\log_2(\snr_{DF1})}+\frac{1}{\log_2(\snr_{DF2})}<\frac{1}{\log_2(\snr_{P2P})},
\end{align}
then DF has a lower latency than P2P. The statement of this proposition can be obtained by neglecting $\ln(2)$ from the upper bound on $N_{DF}$ at high $\snr$. This statement is interesting especially in light of \eqref{RateDF} and \eqref{RateP2P}, since DF achieves higher information-theoretic rates than P2P if 
\begin{align*}
\max\left\{\frac{1}{\log_2(\snr_{DF1})},\frac{1}{\log_2(\snr_{DF2})}\right\}<\frac{1}{\log_2(\snr_{P2P})}.
\end{align*}
This condition indicates that the smallest among $\snr_{DF1}$ and $\snr_{DF2}$ decides whether or not DF performs better than P2P in the information-theoretic sense ($B\to\infty$). However, for finite transmission $B<\infty$, both channels matter. 

A similar comparison between the latencies of AF and P2P leads to the statement of Proposition \ref{Prop:AF}. At high $\snr$, if 
\begin{align}
\label{AFCond1}
\frac{2}{\log_2(\snr_{AF})}<\frac{1}{\log_2(\snr_{P2P})},
\end{align}
then AF has a lower latency than P2P. Similar to the discussion on DF above, at high $\snr$, AF achieves higher information-theoretic  rate than P2P if $\frac{1}{\log_2(\snr_{AF})}<\frac{1}{\log_2(\snr_{P2P})}$ but achieves lower latency only if condition \eqref{AFCond1} holds, which is stricter.

\section{Conclusions}
We have derived the transmission latency of Gaussian transmission using DF and AF in a relay channel and compared that with the latency of the P2P channel scheme which does not use the relay. It turns out that the relay reduces the latency of the transmission under a condition on the $\snr$'s. The main insights from this paper are two fold. First, relaying can be used to decrease the latency of a transmission, but the relay has to be properly set-up. Second, the information-theoretic achievable rate of a scheme is not suitable for analysing the latency of a scheme for communicating a given number of bits under a reliability requirement, since information-theoretic rates are tailored for infinite transmission. As an extension, it would be interesting to examine the impact of relays on the latency in a fading channel, in RC's with multiple relays, and in half-duplex RC's.

\bibliography{myBib}

\end{document}